\documentclass[twocolumn]{aastex6}

\usepackage{tabularx}
\usepackage{epsfig}
\usepackage[english]{babel}
\usepackage[utf8]{inputenc}
\usepackage{amsmath,amsfonts}
\usepackage{url}
\usepackage{booktabs}
\usepackage{comment}
\usepackage{multirow}

\usepackage[titletoc]{appendix}
\usepackage{natbib}
\usepackage{ulem}

\begin{document}

\title{Accurate orbital solution for the new and metal poor eclipsing binary Tycho~5227-1023-1}

\author{
G. Traven \altaffilmark{1},
U. Munari \altaffilmark{2},
S. Dallaporta \altaffilmark{3},
T. Zwitter \altaffilmark{1}
}

\altaffiltext{1}{Faculty of Mathematics and Physics, University of Ljubljana, Jadranska 19, 1000 Ljubljana, Slovenia; \\ e-mail: gregor.traven@fmf.uni-lj.si}
\altaffiltext{2}{INAF Osservatorio Astronomico di Padova, 36012, Asiago, VI, Italy}
\altaffiltext{3}{ANS Collaboration, c/o Astronomical Observatory, 36012 Asiago (VI), Italy}

\begin{abstract}
The orbit and physical parameters of the previously unsolved double-lined eclipsing binary Tyc 5227-1023-1, discovered during the search for RR Lyr variables candidate members of the Aquarius stream, are derived using high resolution \'echelle spectroscopy and $V$, $i^\prime$ photometry. Synthetic spectral analysis of both components has been performed, yielding metallicity [M/H] = $-0.63 \pm 0.11$ for both stars and the temperature of the secondary which is in close agreement with the one from the orbital solution, while the temperature of the primary is determined from photometry (T$_1 = 6350$ K). The masses and radii ($M_1 = 0.96 \pm 0.02, M_2 = 0.84 \pm 0.01$ M$_{\odot}$, $R_1 = 1.39 \pm 0.01, R_2 = 0.98 \pm 0.01$ R$_{\odot}$) reveal that both stars have already slightly evolved away from the Main Sequence band having an age of about $7$ Gyr, and the results of synthetic spectral analysis support the claim of co-rotation with the orbital motion. The radial velocity of the system is $-60 \pm 2$ km sec$^{-1}$ while its distance, computed from orbital parameters and the derived reddening $E_{B-V}$ = 0.053, is $496 \pm 35$ pc. Even though Tyc 5227-1023-1 was initially treated as a possible member of the Aquarius stream, the results presented here disagree with reported values for this ancient structure, and suggest a likely membership of the thick disk.
\end{abstract}

\maketitle

\section{Introduction}

Stars are the main baryonic building blocks of galaxies, and since multiplicity is known to be common in the general stellar population \citep{1991AA...248..485D, 2010ApJS..190....1R, 1999ApJ...521..682A, 2009AA...493..947S, 2014LRR....17....3P}, the wealth of information provided by multiple and especially binary systems contributes heavily to our understanding of galactic structure and evolution. Multiple systems not only play a key astrophysical role over the whole HR diagram, but facilitate determination of all important physical parameters such as masses, radii, temperatures, luminosities, and distance \citep{2010AARv..18...67T, 2011AA...525A...2B, 2012AA...543A.106B}.  

The unique geometrical properties of double-lined eclipsing binaries (SB2 EBs) make them forerunners in the quest for obtaining accurate fundamental properties of stars, first of all masses and radii, using a minimum of theoretical assumptions and modelling \citep{2004AA...418L..31M, 2010AARv..18...67T}. Given that the component stars have the same age and initial chemical composition, eclipsing binaries represent a formidable benchmark for the validation of the current generation of stellar evolutionary models (\cite{1991AARv...3...91A, 2002AA...396..551L} and references therein).

The photometric variability of eclipsing binary Tyc 5227-1023-1 (RA: 22 00 52.6 DEC: $-$03 42 12.4, J2000) was first noted by \cite{2014JAD....20....4M} who reported about 180 new field variables discovered as a by-product of the search for RR Lyr variables candidate members of the Aquarius stream \citep{2014NewA...27....1M}.

The RAVE Survey enabled the discovery of the Aquarius stream by \cite{2011ApJ...728..102W}, describing a chemically coherent structure which originates from the tidal disruption of a 12 Gyr old [$\alpha$/Fe]-enhanced globular cluster of low metallicity ([Fe/H] = $-$1.0; \citealp{2012ApJ...755...35W}). The stream appears to be on a trajectory toward the solar neighbourhood from the direction of Aquarius. The velocity of the infalling stream members increases as they reach the disk of the Galaxy, from $-$160 km sec$^{-1}$ to $-$210 km sec$^{-1}$, with smaller velocities pertaining to the most distant known members at about 3 kpc, and larger ones for those closer to us at about 1 kpc. However, fitting the stellar parameters of stream members to isochrones produces only a crude estimate for their distance. It is therefore important to identify more members and derive robust distances to them, thus enabling the reconstruction of the stream's Galactic orbit and 3-dimensional shape, along with constraining the Galactic gravitational potential in the solar vicinity.

The scarce epoch photometry available to \cite{2014JAD....20....4M} did not allow to classify the type of variability exhibited by Tyc 5227-1023-1. We initally acquired some additional photometric data that suggested it to be an eclipsing binary, and an exploratory high resolution optical spectrum proved it to be a double-lined binary of low metallicity. This immediately boosted our interest in the object, and a full-scale observing campaign was initiated aiming to obtain an accurate orbital solution and therefore a geometrical distance to Tyc 5227-1023-1. Accurate orbital solution for high Galactic latitude, metal poor binaries are rare, and this alone could justify the present investigation. Should our target turn out to be a member of the Aquarius stream, this would further boost the interest in it.

\begin{table}
\caption{CCD photometry in the Landolt's $V$ and SLOAN $i^\prime$ bands of Tyc 5227-1023-1. The columns give the heliocentric JD (\mbox{$-2456000$}), the orbital phase and the magnitudes with their uncertainties. Table \ref{LC} is published in its entirety only electronically, and a portion is shown here for guidance regarding its form and content.}
\footnotesize
\centering
\setlength{\tabcolsep}{4pt}
\begin{tabular}{ cccccc }
\hline\hline

\multicolumn{1}{c}{HJD} & \multicolumn{1}{c}{Phase} & \multicolumn{1}{c}{$V$} & \multicolumn{1}{c}{$err$} &  \multicolumn{1}{c}{$i^\prime$} & \multicolumn{1}{c}{$err$}\\
\hline
841.560 & $-0.5657$ & 11.920 & 0.012 & 11.655 & 0.009 \\
842.532 & $-0.3400$ & 11.924 & 0.006 & 11.649 & 0.005 \\
842.574 & $-0.3302$ & 11.924 & 0.003 & 11.643 & 0.005 \\
850.545 & $-0.4792$ & 11.974 & 0.008 & 11.693 & 0.007 \\
857.489 & $-0.8665$ & 11.933 & 0.004 & 11.660 & 0.004 \\
...\\
\hline
\end{tabular}
\label{LC}  
\end{table}

\section{Observational data}

\subsection{Photometry}

CCD photometry in the Landolt's $V$ and SLOAN $i^\prime$ bands of Tyc 5227-1023-1 has been obtained with ANS Collaboration telescope N. 36, which is a 0.30-m Ritchey-Cretien telescope located in Cembra (Trento, Italy).  It is equipped with an SBIG ST-8 CCD camera, 1530$\times$1020 array, 9~$\mu$m pixels $\equiv$ 0.77$^{\prime\prime}$/pix, with a field of view of 19$^\prime \times 13^\prime$. The $V$ and $i^\prime$ filters are from Schuler and Astrodon, respectively. The data are given in \ref{LC} (available in full only electronically), where the quoted uncertainties are the total error budget, combining quadratically the measurement error on the variable with the error of the transformation from the instantaneous local photometric system to the standard one, as defined by the local photometric sequence extracted from the APASS survey \citep{2012JAVSO..40..430H,2014JAD....20....4M} which is calibrated against the \cite{2009AJ....137.4186L} and \cite{2002AJ....123.2121S} equatorial standards. Technical details of the ANS Collaboration network of telescopes running since 2005, their operational procedures and sample results are presented by \citep{2012BaltA..21...13M}. Detailed analysis of the photometric performances and measurements of the actual transmission profiles for all the photometric filter sets in use is presented by \citep{2012BaltA..21...22M}. All measurements on Tyc 5227-1023-1 were carried out with aperture photometry, the long focal length of the telescope and the absence of nearby contaminating stars not requiring the use of PSF fitting.

\begin{figure*}[!htp]
   \centering
   \includegraphics[trim = 0mm 15mm 0mm 25mm, clip, width=0.85\linewidth]{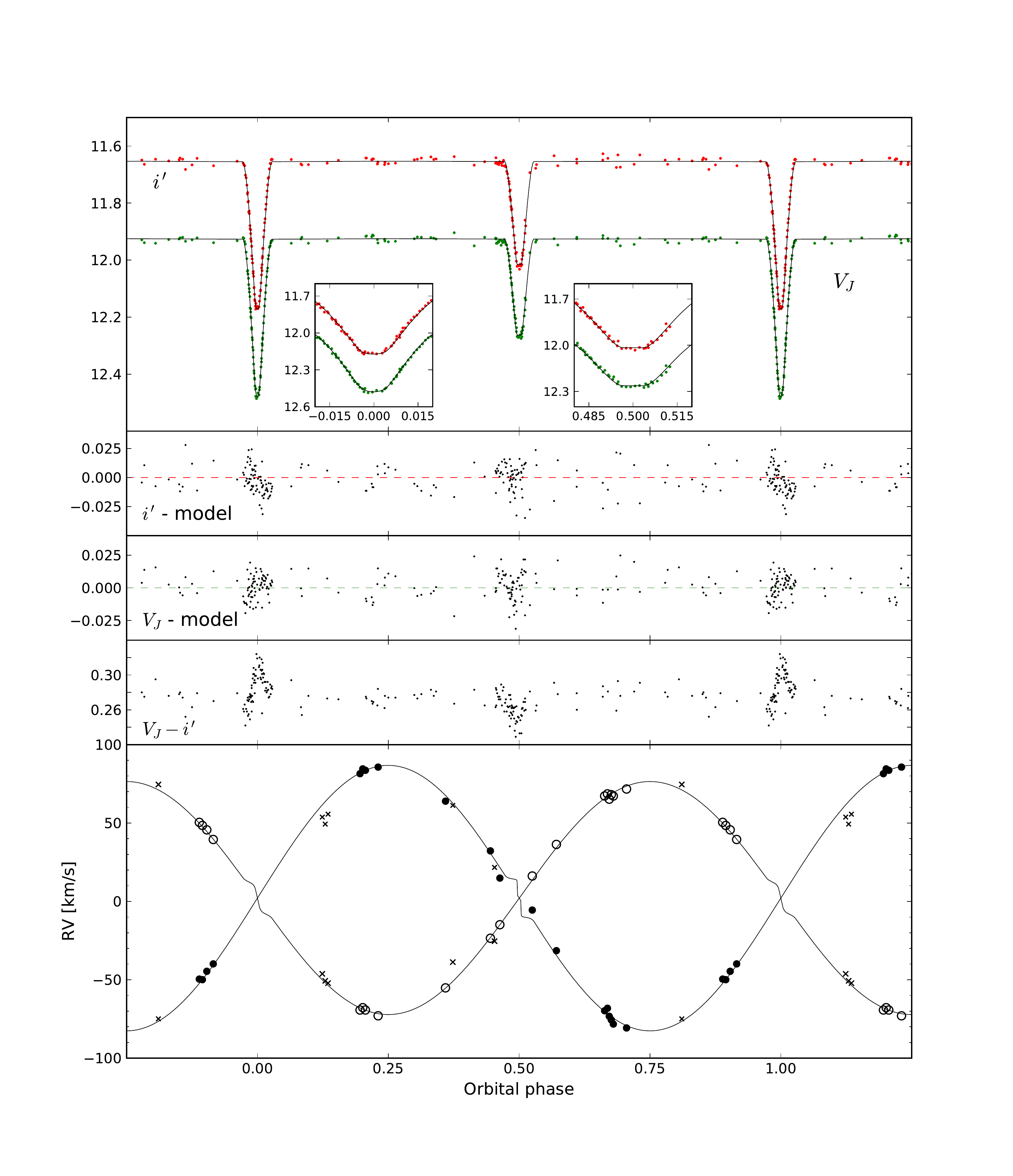}                        
   \caption{Our $V_J$, $i^\prime$, $i^\prime$-model, $V_J$-model, $V_J$ - $i^\prime$ and radial velocity data of Tyc 5227-1023-1. In the radial velocity panel, the open circles indicate the hotter and more massive star 1, while the filled circles pertain to the cooler and less massive star 2. The RV values for PH$_0$ spectrum are not plotted, while RV values rejected from the solution (see Section \ref{orbi}) are marked by x's. The orbital solution from Table \ref{PHOEBE} is over-plotted to the data and used in the second and third panel from the top.}
   \label{sol}
\end{figure*}

\subsection{Spectroscopy}

Spectra of Tyc 5227-1023-1 have been secured in 2015--2016 with the \'echelle+CCD spectrograph mounted on the 1.82 m telescope operated by Osservatorio Astronomico di Padova atop Mt. Ekar (Asiago). The instrumentation and observing set-up match those described by \cite{2004AA...417.1083S}, to which we refer for details of the observing mode. Here we recall that the $3600-7400$ \AA\, wavelength region is covered in 30 orders at a resolving power of 20,000. A journal of the observations is given in Table \ref{RV} for 26 obtained spectra with exposure times of 1200, 1800, and one of 2700 sec, which provide a moderate S/N ratio while avoiding smearing due to the orbital motion (1500 sec correspond to less than 2\% in orbital period). The first spectrum listed in the table was recorded practically at zero phase (hereafter PH$_0$ spectrum), and was used to measure the systemic velocity as well as a template for RV measurements.

\begin{table}
\caption{Heliocentric radial velocities (km sec$^{-1}$) of Tyc 5227-1023-1. The columns give the spectrum number (from the Asiago \'echelle log book), the heliocentric JD (\mbox{$-2457000$}), the orbital phase, exposure time (sec), the radial velocities of the two components in km sec$^{-1}$ and the corresponding uncertainties. The last column gives the S/N of the spectrum (per pixel) averaged over the wavelength range considered in the radial velocity analysis. Listed in the last six rows are RV values which are rejected from the solution (see Section \ref{orbi}).}
\footnotesize
\centering
\setlength{\tabcolsep}{4pt}
\begin{tabular}{ l l l l r l r l r }
\hline\hline
\multicolumn{1}{c}{\#} & \multicolumn{1}{c}{HJD} & \multicolumn{1}{c}{Phase} & \multicolumn{1}{c}{t} & \multicolumn{2}{c}{Star 1} & \multicolumn{2}{c}{Star 2} & \multicolumn{1}{c}{{\scriptsize $<$S/N$>$}}\\
\cmidrule(lr){5-6}\cmidrule(lr){7-8}
 & & & & RV & $err$ & RV & $err$ & \\
\hline
56808 & 326.289 & 0.9999 & 1800 & & & & & 23 \\
56836 & 327.282 & 0.2305 & 1800 & -73.0 & 1.6 & 85.7 & 2.0 & 31 \\
56891 & 328.284 & 0.4631 & 1800 & -14.9 & 1.9 & 14.9 & 4.8 & 28 \\
56953 & 329.326 & 0.7052 & 1800 & 71.7 & 1.0 & -80.7 & 1.0 & 31 \\
56986 & 330.232 & 0.9155 & 1800 & 39.5 & 1.6 & -39.9 & 2.1 & 32 \\
57011 & 350.280 & 0.5711 & 1800 & 36.3 & 4.1 & -31.4 & 6.1 & 22 \\
57277 & 380.225 & 0.5251 & 1200 & 16.2 & 2.0 & -5.5 & 2.1 & 20 \\
58030 & 586.577 & 0.4450 & 1800 & -23.5 & 1.8 & 32.2 & 5.1 & 22 \\
58053 & 587.555 & 0.6721 & 1200 & 65.2 & 2.3 & -73.3 & 2.1 & 16 \\
58055 & 587.572 & 0.6760 & 1200 & 68.1 & 1.5 & -75.5 & 3.0 & 17 \\
58057 & 587.589 & 0.6799 & 1200 & 67.2 & 1.4 & -78.3 & 3.5 & 18 \\
58074 & 588.550 & 0.9032 & 2700 & 45.7 & 1.3 & -44.6 & 2.5 & 31 \\
58592 & 736.222 & 0.1959 & 1800 & -69.3 & 1.5 & 81.5 & 1.9 & 18 \\
58593 & 736.244 & 0.2011 & 1800 & -67.8 & 2.0 & 84.5 & 3.8 & 23 \\
58594 & 736.266 & 0.2062 & 1800 & -69.3 & 1.5 & 83.6 & 2.8 & 24 \\
58666 & 738.234 & 0.6633 & 1800 & 67.3 & 1.9 & -69.7 & 2.8 & 26 \\
58667 & 738.257 & 0.6687 & 1800 & 68.6 & 1.1 & -68.2 & 1.7 & 26 \\
58710 & 739.205 & 0.8889 & 1800 & 50.4 & 2.5 & -49.6 & 2.5 & 24 \\
58712 & 739.231 & 0.8948 & 1800 & 48.4 & 1.5 & -49.9 & 2.1 & 26 \\
58799 & 741.231 & 0.3592 & 1800 & -55.2 & 1.9 & 63.9 & 11.1 & 19 \\
\hline
57068 & 351.312 & 0.8107 & 1800 & 74.6 & 1.6 & -74.9 & 1.8 & 18 \\
57337 & 384.221 & 0.4530 & 1800 & -25.4 & 2.1 & 21.7 & 3.4 & 25 \\
58120 & 590.574 & 0.3732 & 3600 & -38.7 & 2.6 & 61.2 & 2.6 & 24 \\
58753 & 740.218 & 0.1239 & 1800 & -46.2 & 1.8 & 53.8 & 1.8 & 23 \\
58755 & 740.242 & 0.1295 & 1800 & -50.7 & 1.5 & 49.3 & 1.5 & 25 \\
58757 & 740.265 & 0.1349 & 1800 & -52.1 & 1.5 & 55.6 & 8.5 & 22 \\
\hline
\end{tabular}
\label{RV}  
\end{table}

The spectra have been extracted and calibrated in a standard fashion with IRAF. The wavelength solution has been derived simultaneously for all 30 recorded \'echelle orders, with an average r.m.s of 0.32 km sec$^{-1}$.

\subsection{Systemic velocity}

The systemic velocity of Tyc 5227-1023-1 is measured on the PH$_0$ spectrum. The IRAF's fxcor routine is applied to 22 \'echelle orders \#34-55 [$4000-6700$ \AA], using a similar synthetic spectrum selected from the \cite{2005AA...442.1127M} synthetic spectral atlas computed at the same 20~000 resolving power as the \'echelle scientific spectra. The average value of the systemic radial velocity is $-62.47$ km sec$^{-1}$ with the uncertainty of 2 km sec$^{-1}$ (cf. Table \ref{PHOEBE}). This value is later adjusted by the orbital solution (see Section \ref{paras}).

\subsection{Radial velocities} 

We use TODCOR, a two-dimensional cross-correlation algorithm \citep{1994ApJ...420..806Z}, to derive the radial velocities. This technique obtains Doppler shifts of both stellar components simultaneously by employing a multiple correlation approach, producing at the same time their intensity ratio. We apply it to the 22 \'echelle orders \#34-55 covering the $4000-6700$ \AA\, range. Each order is trimmed 25\% at both ends, corresponding to a redundant region where adjacent orders overlap. The lower instrument response in these end regions leads to: ({\textit a}) degradation of wavelength solution accuracy, ({\textit b}) poorer S/N, and ({\textit c}) a steeper continuum since normalisation to unity is more difficult. Retaining only the central 50\% of each order alleviates these issues. 

The PH$_0$ spectrum was used as the template for TODCOR in deriving radial velocities. The reason for using a larger number of orders than what is detailed in \cite{2004AA...417.1083S} is the lower S/N of spectra in this study, resulting in some of the solutions from arbitrary orders being invalid (e.g. both radial velocities being either larger or smaller than the systemic velocity). The radial velocity values from individual orders are also rejected if they lie outside the 2.5 $\sigma$ of their distribution, which removes the remaining anomalies. On average, results from 12 orders out of 22 are kept, and the final values of radial velocities are summarized in Table \ref{RV}. The mean uncertainty of values used in the orbital solution (see Section \ref{orbi}) is 3.32 km sec$^{-1}$ for star 2 (the fainter of the two), and 1.81 km sec$^{-1}$ for star 1. The small variations in the temperature of the template do not have a significant effect on TODCOR results.

\section{Orbital solution} \label{orbi}

A simultaneous spectroscopic and photometric solution for Tyc 5227-1023-1 was obtained with the PHOEBE 0.31 code \citep{2005ApJ...628..426P, 2005ApSS.296..315P} which is based on the models of \cite{1971ApJ...166..605W} and \cite{1979ApJ...234.1054W}. Orbital modelling was performed using the \textit{detached binary} system option in PHOEBE, appropriate for our case. Solution of the binary system is obtained following the Bayesian approach to parameter estimation, employing a Markov chain Monte Carlo (MCMC) Ensemble sampler \textit{emcee} \citep{2013PASP..125..306F}. At each step in the surveyed parameter space, a PHOEBE orbital solution is computed, with the initial starting point randomly determined from uniform priors. The effective temperature of star 1 and the metallicity for both stars are determined and fixed in an iterative procedure of combining results from: (1) the orbital solution, (2) available photometric data, and (3) atmospheric analysis (see Section \ref{temp}, \ref{atmo}, and Table \ref{PHOEBE}), while the remaining parameters of the binary system are adjusted. A logarithmic law for limb darkening is assumed using the native PHOEBE limb darkening tables for the appropriate T$_{\mathrm{eff}}$, log g, and [M/H], computed per-passband. RV measurements outside the 2.5 $\sigma$ of their distribution around the solution are iteratively rejected and are listed as the last six rows of Table \ref{RV} and marked in Fig. \ref{sol}. In row order, the first five of these are rejected after the first orbital solution converges, the last one after the second solution, and none are rejected after the third (final) solution. The solutions converged after $\sim$ 7000 MCMC iterations employing 256 walkers. For check and completeness, we re-run full orbital solutions also with linear and square-root limb darkening laws as well as for metallicities [M/H]=[M/H]$_{\mathrm{fixed}} \pm 0.2$. The response of orbital solution to these changes was minimal, with orbital parameters not varying by more than their uncertainties. 

\begin{table}
\caption{Orbital solution (over-plotted to observed data in Fig. \ref{sol}) and atmospheric parameters from the $\chi^2$ fit to synthetic spectra for Tyc 5227-1023-1. For parameters derived by PHOEBE ({\sl Orbital solution}), uncertainties from posterior distributions are reported (see Figure \ref{post} in Appendix), whereas for others, as well as for the distance to the system, we state the most reliable and propagated uncertainty estimates. $T_1$ is derived from photometry and fixed in the orbital solution which adjusts only the difference $T_1 - T_2$ (in this case the ratio of both temperatures). Propagation of the uncertainty of $T_1$ onto other derived properties and general remarks on results in this table are discussed in Section \ref{disc}.}
\footnotesize
\centering
\setlength{\tabcolsep}{4pt}
\begin{tabular}{ llrl }
\hline\hline
\multicolumn{4}{c}{Orbital solution} \\
\\
{\sl P} & (d) & 4.306192 & $^{+0.000004}_{-0.000004}$ \\[1.3ex] 
{\sl T$_{\rm 0}$} & (HJD) & 2457003.3261 & $^{+0.0002}_{-0.0002}$ \\[1.3ex] 
{\sl $K_1$} & (km sec$^{-1}$) & 74.4 & $^{+0.5}_{-0.5}$ \\[1.3ex] 
{\sl $K_2$} & (km sec$^{-1}$) & 84.7 & $^{+0.7}_{-0.7}$ \\[1.3ex] 
{\sl a} & (R$_{\odot }$) & 13.5379 & $^{+0.0681}_{-0.0644}$ \\[1.3ex] 
{\sl V$_{\gamma }$} & (km sec$^{-1}$) & 2.12 & $^{+0.34}_{-0.34}$ \\[1.3ex] 
{\sl q = $\frac{m_2}{m_1}$} & (deg) & 0.8777 & $^{+0.0094}_{-0.0091}$ \\[1.3ex] 
{\sl i} &  & 88.87 & $^{+0.18}_{-0.13}$ \\[1.3ex] 
{\sl e} &  & 0.00029 & $^{+0.00013}_{-0.00012}$ \\[1.3ex] 
{\sl T$_{1}$ - T$_{2}$} & (K) & 403 & $^{+7}_{-7}$ \\[1.3ex] 
{\sl $\Omega_{1}$} &  & 10.63 & $^{+0.05}_{-0.04}$ \\[1.3ex] 
{\sl $\Omega_{2}$} &  & 13.23 & $^{+0.16}_{-0.16}$ \\[1.3ex] 
{$r_1$} & ($R_1$/{\sl a}) & 0.1026 & $^{+0.0005}_{-0.0005}$ \\[1.3ex] 
{$r_2$} & ($R_2$/{\sl a}) & 0.0721 & $^{+0.0007}_{-0.0006}$ \\[1.3ex] 
{$R_1$} & (R$_\odot$) & 1.388 & $^{+0.010}_{-0.010}$ \\[1.3ex] 
{$R_2$} & (R$_\odot$) & 0.977 & $^{+0.011}_{-0.009}$ \\[1.3ex] 
{$M_1$} & (M$_\odot$) & 0.9560 & $^{+0.0167}_{-0.0155}$ \\[1.3ex] 
{$M_2$} & (M$_\odot$) & 0.8391 & $^{+0.0123}_{-0.0114}$ \\[1.3ex] 
{M$_{bol, 1}$} &  & 3.61 & $^{+0.02}_{-0.02}$ \\[1.3ex] 
{M$_{bol, 2}$} &  & 4.66 & $^{+0.02}_{-0.02}$ \\[1.3ex] 
{log $g_1$} & (cgs) & 4.13 & $^{+0.01}_{-0.01}$ \\[1.3ex] 
{log $g_2$} & (cgs) & 4.38 & $^{+0.01}_{-0.01}$ \\[1.3ex] 
{distance} & (pc) & 496 & $\pm$ 35 \\[1.3ex] 
\\
\multicolumn{4}{c}{Photometric temperature} \\
 & & & \\
{\sl T$_1$} & (K) & 6350 & $\pm$ 200\\
\\
\multicolumn{4}{c}{Atmospheric analysis} \\
 & & & \\
{\sl T$_2$} & (K) & 5923 & $\pm$ 213 \\
\text{[M/H]} & &  -0.63 & $\pm$ 0.11\\
\\
\multicolumn{4}{c}{Systemic velocity} \\
 & & & \\
$V_{\mathrm{sys}}$ & (km sec$^{-1}$) & -60.35 & $\pm$ 3\\
\hline
\end{tabular}
\label{PHOEBE}  
\end{table}

\subsection{Effective temperature of the primary star} \label{temp}

Initial estimate of the spectral type of Tyc 5227-1023-1 was done by visual comparison of a low resolution spectrum to the spectral classification standard stars and evaluated to G2V (5860 K) -- G3V (5770 K). The temperatures are taken from \cite{2004AJ....128..829B} ATLAS results for dwarfs and the low resolution spectrum was obtained with the Asiago \mbox{1.22m} telescope + B\&C spectrograph. However, considering the low metallicity which could have affected the spectral classification in this part of the HR diagram, 
we prefer to rely on the $B-V$ colour index, for which high quality direct T$_{\mathrm{eff}}$ calibrations exist while they are missing for $V-i^\prime$. 

We obtain the T$_{\mathrm{eff}}$ for star 1 with the transformation of Tycho photometry to the Johnson system following \cite{2000PASP..112..961B}. The uncertainties of the Tycho $B_T$ and $V_T$ are $0.32$ and $0.27$ \citep{2000AA...355L..27H}, indicating the dispersion of the measures due to Tyc 5227-1023-1 being a variable with amplitude $> 0.3$ mag, however this does not reflect into a wrong mean value\footnote{We have checked the soundness of $B_T-V_T = 0.59$ reported in the Tycho catalogue by imaging the field around Tyc 5227-1023-1 (observation outside of eclipse), including other Tycho stars for comparison, whereby deriving a value of $B_T-V_T = 0.55 \pm 0.06$ which is in good agreement with the one from the catalogue.}. The reddening from Section \ref{red} is used to derive $(B-V)_0 = (B-V) - E_{B-V} = 0.55 - 0.053 = 0.497$. With star 1 slightly evolved from the Main Sequence (see Table \ref{PHOEBE} and Figure \ref{theo}), this translates to $T_1 = 6350$ and the spectral type of F7IV-V, following \cite{1970AA.....4..234F} and \cite{2004AJ....128..829B}. Taking into account the colour-T$_{\mathrm{eff}}$ relation, its dependence on metallicity, reddening, and photometric system transformations, the uncertainty on $T_1$ is evaluated to $200$ K. The excellent agreement of the derived temperatures with the position of the stars on the isochrones and evolutionary tracks in Figure \ref{theo} confirmed us on the choice of $T_1$.

\subsection{Atmospheric analysis} \label{atmo}

Considering the moderate S/N of our \'echelle spectra, it is not feasible to resolve the degeneracy among the stellar parameters by means of synthetic spectral fitting. Nevertheless, we use a straightforward atmospheric analysis, performing a simultaneous $\chi^2$ fitting of both components together with the constraints from the orbital solution to derive the metallicity of both stars and to check their rotational velocities for orbital/rotational synchronisation. This is done on the two most appropriate scientific exposures close to quadrature (56953 and 56836 in Table \ref{RV}), using the synthetic atlas of spectra \citep{2005AA...442.1127M}. For a more reliable convergence, only six adjacent \'echelle orders (\#40-45) that cover the wavelength range 4890-5690 \AA\, were selected for their position close to the optical axis of the spectrograph where optical quality is the best and the S/N reaches peak values. Before fitting, each order was trimmed so as to retain only the central 25\%, where the instrumental response and PSF sharpness are the best, providing a measured average resolving power of 18,000.  

The temperature of star 1 is derived from photometry as described in Section \ref{temp}, while the appropriate grid of temperatures for star 2 and metallicities, obtained with linear interpolation, was chosen (T$_{\mathrm{step}} = 10$ K, [M/H]$_{\mathrm{step}} = 0.05$). The low reddening and the similarity of the two stars allow us to set the luminosity ratio (L$_2$/L$_1$ = 0.38) and the surface gravities as given by the orbital solution. Fixing the synchronised rotational velocities based on the orbital solution always produces a better fit. The results of atmospheric analysis therefore support the claim of rotational synchronisation and provide metallicity as the fundamental parameter of the system along with the temperature of star 2.

\subsection{Reddening and distance} \label{red}
To derive the photometric temperature of the system and compute the distance to Tyc 5227-1023-1 from the orbital solution, reddening is evaluated by adopting the statistical 3D approach of \cite{2014AJ....148...81M} that works particularly well for the region of Aquarius. This model is essentially based on a homogeneous slab of dust extending for 140 pc on either side of the Galactic plane, causing a reddening of $E_{B-V}$ = 0.036 at the poles. We obtain the first distance estimate by neglecting the reddening and using only a bolometric correction for star 1 (BC$_1$ = $-0.006$) and star 2 (BC$_2$ = $-0.052$). After confirming that our system is well beyond the region of dust in our model, a color excess $E_{B-V} = 0.053$ is determined and used in the standard A$_V$ = $E_{B-V} \times 3.1$ reddening law. The bolometric correction is calculated in iteration with the derived temperatures of both stars employing the prescription from \cite{2010AJ....140.1158T}. Using the above values, the final distance is computed together with the rest of the parameters in the orbital solution with PHOEBE.

\subsection{Physical parameters} \label{paras}
Final values of parameters with most reliable uncertainty estimates for Tyc 5227-1023-1 binary system are given in Table \ref{PHOEBE}. Although the systemic velocity $V_{\mathrm{sys}}$ is measured on the PH$_0$ spectrum, used as a template for TODCOR to derive radial velocities of other spectra, the $V_{\gamma}$ parameter was nevertheless left free to be adjusted due to the uncertainty in $V_{\mathrm{sys}}$ determination. In the final solution, $V_{\mathrm{sys}}$ is corrected for the value of $V_{\gamma}$. 

To sample the convergence and achieved minimum of the $\chi^2$ fit in atmospheric analysis, we retain the best 50 results, yielding the temperature of star 2 (5923 $\pm 73$ K) and metallicity (-0.63 $\pm 0.02$). All system parameters are well constrained by the orbital solution in Table \ref{PHOEBE}, with formal accuracies of 1.7 and 1.5\% on the masses, and 0.8 and 1.1\% on the radii. The synchronized rotational velocities of the two stars would be 16.3 km sec$^{-1}$ and 11.5 km sec$^{-1}$. The temperature of star 2 ($T_2 = 5947$ K) is in good agreement with that from $\chi^2$ atmospheric fit ($\triangle T_2 = 24$ K).

The posterior distributions of parameters adjusted in the PHOEBE orbital solution with MCMC are plotted in Figure \ref{post} in Appendix. Values of these parameters in Table \ref{PHOEBE} and their uncertainties are determined from the 16th, 50th, and 84th percentile of their distributions.

\begin{figure}[!htp]
   \centering
   \includegraphics[trim = 25mm 0mm 35mm 10mm, clip, width=1\linewidth]{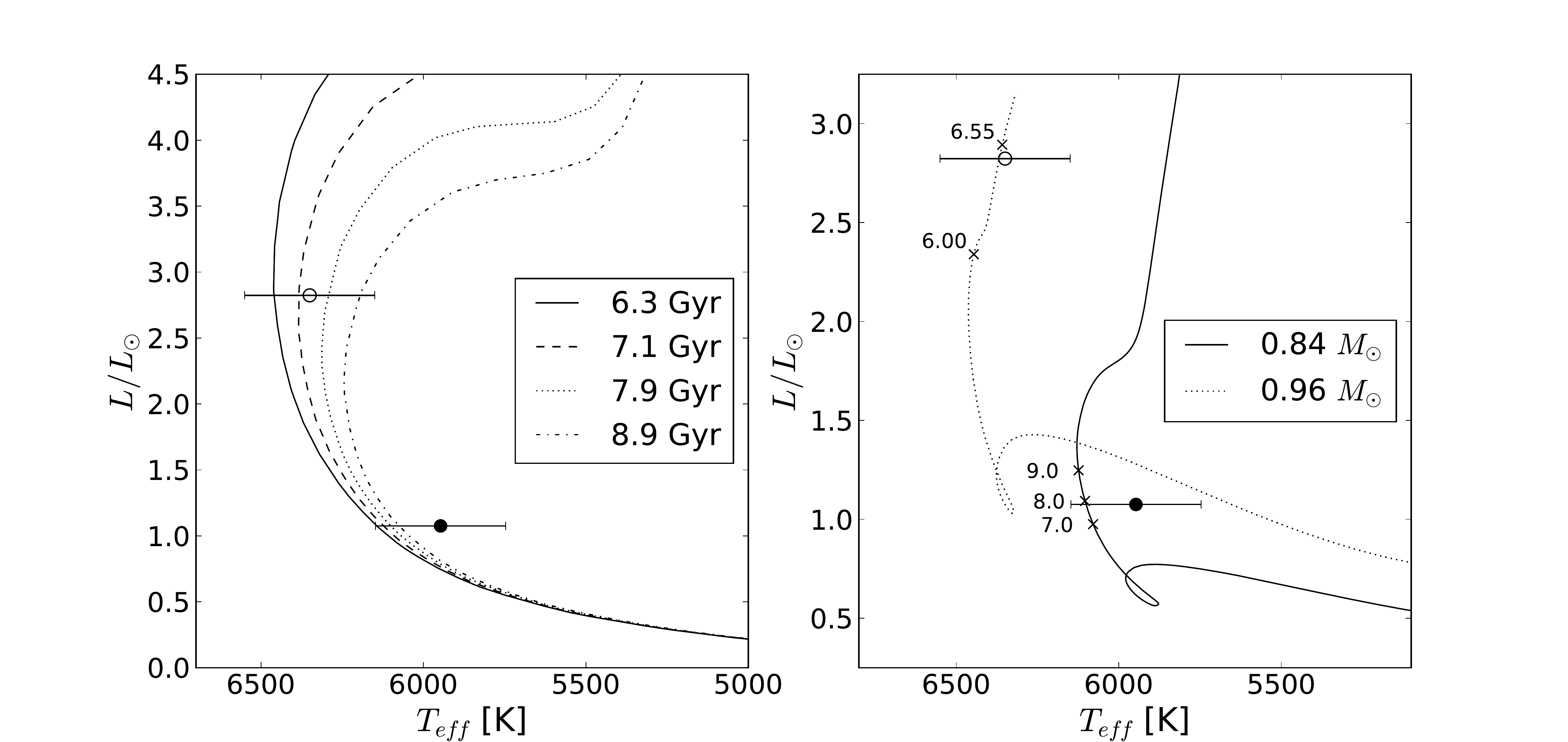}                        
   \caption{Comparison of both stars' temperature and luminosity as given in the orbital solution (cf. Table \ref{PHOEBE}) with those produced by theoretical stellar models. Plots contain Padova theoretical isochrones and evolutionary tracks, which have been derived by interpolating adjacent points from the computed grid ($0.8 - 0.85$ and $0.95 - 1$ $M_{\odot}$, and $0.002$ and $0.004$ for Z). Open circles denote the hotter and more massive star 1, whereas filled circles refer to the cooler and less massive star 2. Age on the evolutionary tracks is marked in Gyr. The error bars on star 1 indicate the uncertainty of photometric evaluation for $T_1$, while those on star 2 are slightly larger due to the additional uncertainty from the orbital solution for $T_2$.}
   \label{theo}
\end{figure}

\subsection{Comparison with theoretical stellar models}

We compare the physical parameters from the orbital solution of Tyc 5227-1023-1 with Padova theoretical models (\cite{2012MNRAS.427..127B,2014MNRAS.445.4287T}, and references therein). The position of the two components on the $L$, $T_{\mathrm{eff}}$ plane is presented in Figure \ref{theo}, where a comparison is provided with isochrones and evolutionary tracks appropriate for the masses (0.96 and 0.84 M$_{\odot}$) and metallicity ([M/H] = log Z/Z$_{\odot}$ = $-0.63$, where Z = 0.0035 and Z$_{\odot}$ = $0.0152$) of the system components, while rotation of both stars is not taken into account. They have been obtained via interpolation over the grid computed by the Padova theoretical group. The agreement between theoretical models and the more reliable solution for star 1 (the primary) predicts the binary's age of about $7$ Gyr.

\section{Discussion} \label{disc}

The physical parameters of Tyc 5227-1023-1 suggest that both components of the binary system have already slightly evolved from the Main Sequence. We are able to derive masses and radii of both stars to formal accuracies of 2\% or better. However, we note that while the formal accuracies of the orbital solution are excellent due to the very high quality photometric data and reasonably accurate radial velocities, there are other factors that have to be taken into account. The comparison of orbital solution to photometry data in Figure \ref{sol} shows a slight asymmetry in the region of the minima, which could be explained by the inclusion of third light in the fit, the atmospheric conditions and calibration of photometric observations at different epohs or by the actual activity of the stars. A possible third body in the system might also contribute to the time variations in the light curve. All these effects can influence the radii determination given in Table \ref{PHOEBE} and increase their uncertainty. We also investigate certain other contributions to the uncertainty on the masses from Table \ref{PHOEBE} by: (1) fixing  the eccentricity to zero, (2) removing all but a few photometry data points from the fit, so as to not have the solution affected or driven primarily by the light curve, and (3) removing the offset to the RV of the system ($V_{\gamma}$ parameter) from the fit, where $V_{\mathrm{sys}}$ is determined from the PH$_0$ spectrum. In the first two cases, the uncertainty on masses determined by MCMC procedure is almost the same, whereas the change in value of both masses is much less than their uncertainty given in Table \ref{PHOEBE}. The third case, however, produces a significant effect of reducing the primary's mass by 1 $\sigma$ compared to the value given in Table \ref{PHOEBE}.

Some of our results are furthermore affected by uncertainties in evaluating the bolometric correction, reddening, and especially the primary's (star 1) effective temperature, which was determined with colour-T$_{\mathrm{eff}}$ relations. In this respect, the uncertainty for the distance given by the PHOEBE orbital solution ($\pm 3$ pc) has to be corrected for the fact that $T_1$ is evaluated with an accuracy of $\pm 200$ K, yielding final uncertainty on the distance to the system of $\pm 35$ pc. The atmospheric analysis (Section \ref{atmo}) is likewise affected by the temperature uncertainty, producing higher metallicity values for increasing temperatures and lower values for decreasing temperatures. The propagated uncertainty on metallicity ($\pm 0.11$) is therefore larger than the one given in Section \ref{atmo}. Taking the uncertainty on both temperatures and metallicity into account, the position of the stars on evolutionary tracks would shift significantly, producing an uncertainty on age of $\pm 1$ Gyr, with a younger age corresponding to lower temperatures and metallicity. However, the comparison to theoretical models becomes much less reliable when scaling these parameters by their uncertainty.

Observationally, the binary system with its visual magnitude $V = 11.86$ is relatively faint for our instrument and would necessitate much longer exposure times to reach higher values of S/N. This was not feasible for several reasons, including the position on the sky at the time of observations. The observed object was often relatively low above horizon, reaching only 40$^{\circ}$ at culmination. Nevertheless, there is good agreement between the atmospheric analysis and the orbital solution by PHOEBE, where the photometrically derived spectral type and the temperature of the primary is suported by theoretical stellar models. Therefore, we are able to present a reliable solution of the system, together with the estimate of its distance at 496 pc, and a conservative age of $7$ Gyr, based on the primary star's position on isochrones and evolutionary tracks, where the 1 mag fainter secondary is expectedly less well constrained. 

There is an indication of alpha enhancement in our spectra, complying with the properties of the Aquarius stream and the thick disk, but it is detected only when extending the synthetic spectral analysis towards bluer wavelengths, where the results for other parameters become less reliable due to lower S/N. Nevertheless, the systemic velocity ($-60$ km sec$^{-1}$), metallicity ($-0.63$), and age ($\approx 7$ Gyr) of Tyc 5227-1023-1 differ from those assigned to members of the Aquarius stream ($-240 <$ RV $< -160$ km sec$^{-1}$, [M/H] = $-1.0$, age $\approx$ 12 Gyr), so despite the possibility of a partnership to this tidally disrupted structure, which was among the initial drivers for this study, the results of the modelling and atmospheric analysis with $\chi^2$ fit in the end disprove it. 

Tyc 5227-1023-1 has consistent chemistry but relatively high velocity with regard to the thick disk (typical velocity dispersion $\sigma_{z} \approx 34$ km sec$^{-1}$, \cite{2014ApJ...793...51S}), whereas it would kinematically comply well with the galactic halo, although being on the metal-rich end of halo stars, reminiscent of the recently discovered metal-rich halo star born in the Galactic disk \citep{2015MNRAS.447.2046H}. In the latter proposition however, the dynamical ejection from the thick disk into the halo does not seem likely due to this system's mass and binary nature.

The first official release of Gaia data (DR1; \citealt{2016arXiv160904172G}) for stars from the Tycho catalogue gives a parallax measurement of $2.11 \pm 0.8$ mas ($473 \pm 180$ pc), which is in very good agreement with our result on the distance to the system ($496 \pm 35$ pc). Using the numerical integrator NEMO \citep{1995ASPC...77..398T,2010ascl.soft10051B} and the results presented in this work complemented by Gaia astrometric solution (pmRA = $47.67 \pm 2.98$ mas yr$^{-1}$, pmDE = $-8.1 \pm 1.44$ mas yr$^{-1}$), we derive a highly eccentric orbit ($e \approx 0.77, i \approx 7^{\circ}$), having the pericenter and apocenter at 1.1 and 8.6 kpc, respectively. For this approximate orbit evaluation, the maximum height above the galactic plane ($\approx 575$ pc) is consistent with the thick disc population.

\acknowledgments
We would like to thank Andrea Frigo for his support in the reduction of photometric observations. We thank G. Cherini for assistance with some of the photometric observations. G.T. acknowledges the financial support from the Slovenian Research Agency (research core funding No. P1-0188). This work has made use of data from the European Space Agency (ESA)
mission {\it Gaia} (\url{http://www.cosmos.esa.int/gaia}), processed by
the {\it Gaia} Data Processing and Analysis Consortium (DPAC,
\url{http://www.cosmos.esa.int/web/gaia/dpac/consortium}). Funding
for the DPAC has been provided by national institutions, in particular
the institutions participating in the {\it Gaia} Multilateral Agreement.

\bibliography{sample}

\appendix

\section*{Posterior distributions of adjusted parameters} \label{posteriors}

\begin{figure*}[!htp]
   \centering
   \includegraphics[width=1\linewidth]{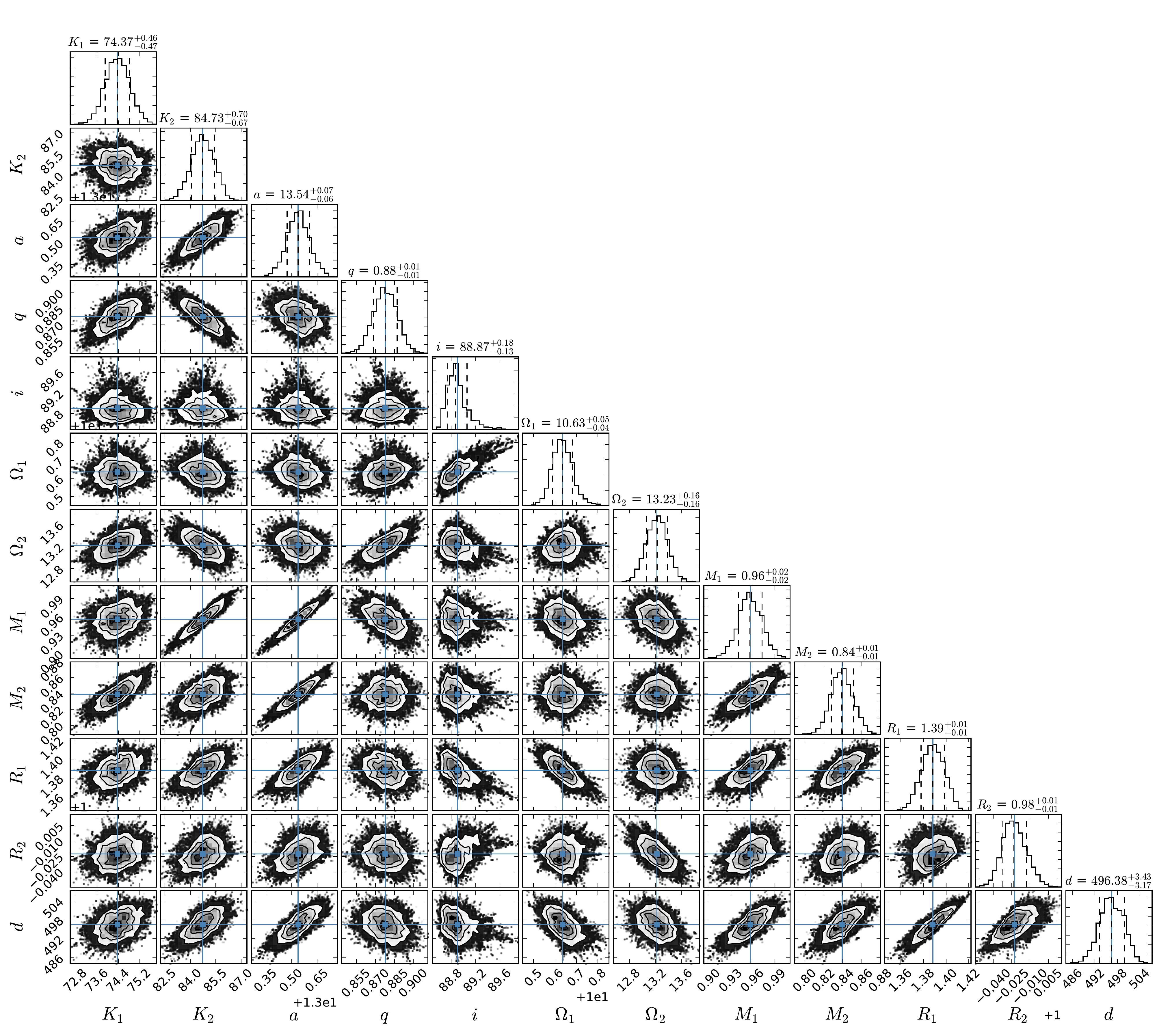}                        
   \caption{The parameter distributions include all values of 256 walkers for the last 1000 iterations after convergence of the orbital solution. The tiles visualize parameter posteriors (diagonal) and parameter correlations (lower left). Dashed lines in posteriors represent the 16th, 50th, and 84th percentile, while the solid lines in all sub-panels mark the mean value.}
   \label{post}
\end{figure*}

Posterior distributions are available for all adjusted parameters from Table \ref{PHOEBE} and a few more that PHOEBE uses internally. Due to their compact representation in the triangle plot and for easier viewing, only the most interesting ones are shown in Figure \ref{post}.

\end{document}